\documentclass[prl,aps,twocolumn,showpacs]{revtex4}

\usepackage{graphicx}
\usepackage{dcolumn}
\usepackage{bm}

\newcommand{\ket}[1]{|#1\rangle}
\newcommand{\bra}[1]{\langle#1|}

\newcommand{\eq}[1]{Eq.~(\ref{#1})}
\newcommand{\eqs}[1]{Eqs.~(\ref{#1})}
\newcommand{\fig}[1]{Fig.\ref{#1}}

\begin{document}

\title{Characterizing quantum phase transitions by single qubit operations}

\author{S. M. Giampaolo, F. Illuminati, and S. De Siena}

\affiliation{Dipartimento di Fisica ``E. R. Caianiello'',
Universit\`a di Salerno, Coherentia CNR-INFM, and INFN Sezione di
Napoli - Gruppo Collegato di Salerno, Via S. Allende, 84081
Baronissi (SA), Italy}


\begin{abstract}

We introduce observable quantities, borrowing from concepts of
quantum information theory, for the characterization of quantum
phase transitions in spin systems. These observables are uniquely
defined in terms of single spin unitary operations. We define the
energy gap between the ground state and the state produced by the
action of a single-qubit local gate. We show that this static
quantity involves only single-site expectations and two-point
correlation functions on the ground state. We then discuss a
dynamical local observable defined as the acceleration of quantum
state evolution after performing an instaneous single-qubit
perturbation on the ground state. This quantity involves
three-point correlations as well. We show that both the static and
the dynamical observables detect and characterize completely
quantum critical points in a class of spin systems.

\end{abstract}

\date{April 7, 2006}

\maketitle

Quantum phase transitions differ from classical ones in that they
are driven by quantum fluctuations. They are induced by the
changes in external control parameters or coupling constants,
which can modify the ground state of a quantum system
\cite{Sachdev}. In analogy with classical phase transitions driven
by thermal fluctuations, quantum phase transitions are usually
analyzed in terms of the scaling behavior of equilibrium
properties and diverging correlation lengths as signatures of
quantum critical points \cite{Sachdev,Hastings}. In the last years
interest in quantum critical phenomena has increased enormously
thanks to several studies that have highlighted some intriguing
connections between quantum phase transitions and the scaling
behavior of nonlocal quantum correlations in the ground state of
complex quantum systems \cite{Osterloh,Osborne,Calabrese}. Quantum
critical points in some classes of quantum spin models can be
characterized by the divergence in the first derivative of the
two-site (two-qubit) concurrence, a proper measure of bipartite
entanglement \cite{Wootters}, with respect to the proper control
parameter, even though recent work on different systems has
revealed that this behavior of the concurrence is not universal at
generic quantum critical points \cite{Roscilde,Gu,Chen},
and different classification schemes have been proposed recently
\cite{Gu2,Wu}.
The studies on the scaling behavior of entanglement prompt
naturally the question on the possible links of entanglement
measures to observable quantities in quantum critical systems.
More generally, as quantum entanglement plays a fundamental role
both in quantum information and quantum computation
\cite{Nielsen}, the further question arises whether it is possible
to reveal structural aspects of quantum critical systems using
other concepts and techniques, beyond entanglement, motivated
and/or imported from the arena of quantum information theory.

Following this line of thought, in this work we discuss the
behavior of  observables associated to unitary transformations on
individual spins, i.e. local unitary single qubit operations. We
will show that they can be used to detect and characterize quantum
critical points in a class of quantum spin systems. First, we will
focus our attention on the energy gap needed to realize single
qubit operations by local excitations above the ground state.
Next, we will consider the short-time diffusion coefficient of the
on site magnetization along a given direction, i.e. the quantum
speed characterizing the dynamics of the magnetization under the
action of single qubit operations. We will first establish that
the observable quantities associated to fundamental local
operations (elementary local gates) can be unambiguously expressed
as functions of one-, two-, and three-point correlation functions
evaluated in the ground state. These relations in turn will allow
to study their behavior when approaching a quantum phase
transition, and to detect and characterize it.

Let us consider the class of spin-$1/2$ ferromagnetic systems
described by the Hamiltonian
\begin{equation}
\label{Hamiltoniana} H = - \frac{J}{2}\sum_{i}
\left[(1+\gamma)\sigma_i^x\sigma_{i+1}^x +
(1-\gamma)\sigma_i^y\sigma_{i+1}^y\right] + h\sum_{i} \sigma_i^z
\, ,
\end{equation}
where $i$ runs over the sites of the lattice, $J$ is the
nearest-neighbor coupling constant, $h$ is the transverse magnetic
field directed along the $z$-axis, $\gamma$ is the anisotropy
parameter, $0 \leq \gamma \leq 1$, $\sigma_i^\alpha$ are the Pauli
matrices ($\alpha$ = $x$, $y$, $z$), and periodic boundary
conditions are assumed throughout. It is convenient to define the
dimensionless reduced coupling constant $\lambda = J/h$. These
models are exactly solvable \cite{Lieb,Barouch,Pfeuty}, and, in
principle, the structure of the correlation functions in the
ground state can be studied accurately. For $\gamma=1$
\eq{Hamiltoniana} reduces to the Ising model whereas for
$\gamma=0$ it reduces to the isotropic $XY$ model. In the interval
$0 < \gamma \le  1 $ the models belong to the Ising universality
class. For $0 < \gamma \le  1 $, in the thermodynamic limit
the system undergoes an order-disorder phase transition
at the critical value $\lambda_c = 1$. If $\lambda
>\lambda_c$ the magnetization $<\sigma_i^x>$
is finite and vanishes at the transition, while the
magnetization $<\sigma_i^z>$ remains finite
for any value of $\lambda$ and vanishes only
in the limit $\lambda \rightarrow \infty$. The case
$\gamma=0$ is in a critical regime for all
$\lambda \ge \lambda_c$.

A local gate (or one qubit gate) is any unitary transformation
acting as the identity $\textbf{1}$ on any subspace defined on any
spin of the lattice except one. Consider the single spin, on which
the local gates act nontrivially, located, say, at a certain site
$k$. Then, the corresponding unitary operator can be written as $U
=\prod_{i \neq k} \textbf{1}_i O_k$, where $O_k$ is a generic
unitary operation acting on the spin placed on site $k$
\cite{Nielsen}. The natural basis in the space of local gates at
lattice site $k$ is formed by the three Pauli matrices
$\sigma_k^x, \, \sigma_k^y, \, \sigma_k^z$, and the identity
$\textbf{1}_k$. For instance, the Hadamard gate is $H_k =
(\sigma_k^x + \sigma_k^z)/\sqrt{2}$, the phase gate is $S_k =
((1-i)\sigma_k^z + (1+i)\textbf{1}_k)/2$, and the spin flip
operator is simply $\sigma_k^x$.

The action of a local gate on the ground state $\ket{g}$ of the
system takes it to a state $\ket{\psi}= O_k \ket{g}$ that, in
general, is not an eigenstate of the many-body Hamiltonian $H$.
Introducing the (dimensionless) energy gap $\Delta E({O}_k) =
(\bra{\psi}H\ket{\psi}-\bra{g}H\ket{g})/h$ it is possible to
determine the exact expressions for the three energy gaps
associated to the fundamental gates in terms of various
correlation functions. We obtain:
\begin{eqnarray}
\label{Energy Gap}
\Delta E_x &=& 2 \lambda (1-\gamma) G_{yy} - 2 M_z  \; , \nonumber \\
\Delta E_y &=& 2 \lambda (1+\gamma) G_{xx} - 2 M_z \; , \\
\Delta E_z &=& 2 \lambda \left[ (1+\gamma) G_{xx}+ (1-\gamma)
G_{yy}  \right] \; , \nonumber
\end{eqnarray}
where $\Delta E_\alpha$ denotes the energy gap due to the action
of the gate $\sigma_k^\alpha$, $M_z$ is the value of the on site
magnetization along the $z$-axis in the ground state: $M_z=\bra{g}
\sigma_k^z \ket{g}$, and $G_{\alpha \alpha}$ denote the values of
the nearest neighbor two-point correlation functions in the ground
state: $G_{\alpha \alpha}=\bra{g} \sigma_k^\alpha
\sigma_{(k+1)}^\alpha\ket{g}$. In the derivation of \eqs{Energy
Gap} we properly exploited the invariance of Hamiltonian
\eq{Hamiltoniana} under translations.

The major significance of the set of \eqs{Energy Gap} is that the
energy gaps between excited states obtained by local gate
operations on the ground state and the ground state itself can be
expressed as functions of (bipartite) expectation values on
$\ket{g}$. Therefore, correlation functions and other ground state
properties can be determined once the energy gaps entering in
\eqs{Energy Gap} are known, and changes in $\ket{g}$ must be
reflected in changes of the energy gaps.

\begin{figure}[t!]
\includegraphics[width=7.5cm]{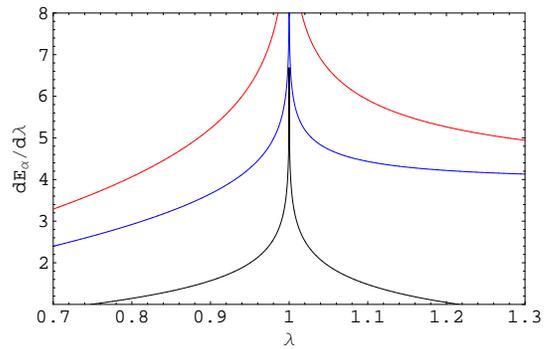}
\caption{First derivatives of the energy gaps as functions of
$\lambda$ in the Ising model ($\gamma=1$). Red line is $d \Delta
E_z /d \lambda$, blue line $d \Delta E_y /d \lambda$, and black
line $- d \Delta E_x/d \lambda$.} \label{f-derivative}
\end{figure}

In \fig{f-derivative} we plot the behavior of the derivatives with
respect to $\lambda$ of the three energy gaps in the case of the
Ising model ($\gamma=1$). Here and in the following of this work
we will exploit the studies performed by Barouch and McCoy
\cite{Barouch} to obtain exact analytical expressions.
We find that the derivatives of the three energy gap diverge
logarithmically at the critical point $\lambda_c =1$ as
\begin{equation}
\label{qptbehaviour} \frac{d \Delta E_\alpha}{d \lambda} =
c_\alpha(\gamma) \ln |\lambda- \lambda_c| + const \; ,
\end{equation}
where, for the Ising model, the values of the coefficients
$c_\alpha(1)$ are $c_z(1)=-1.276$, $c_x(1)=0.638$, and
$c_y(1)=-c_x(1)$.

For $0<\gamma<1$ the models described in \eq{Hamiltoniana} are in
the same universality class of the Ising model. With respect to
the concept of universality - the critical properties depend only
on the dimensionality of the system and on the broken symmetry in
the ordered phase - it is important to verify that the energy gaps
show similar behaviors when the models in this range of values of
$\gamma$ approach the quantum critical point. We have verified the
same property of logarithmic divergence when the anysotropy
parameter varies in the open range $(0,1)$.
For instance, for $\gamma = 0.5$ the derivatives of the energy
gaps close to the quantum critical point obey the same law
\eq{qptbehaviour} with $c_z(0.5)=-2.550$, $c_x(0.5)=0.638$ and
$c_y(0.5)=-c_x(0.5)$. In general, for decreasing $\gamma$, we find
that the value of $c_z$ decreases and, remarkably, that the values
of $c_x$ and $c_y$, and the relation between them, remain
constant.

\begin{figure}[t!]
\includegraphics[width=7.5cm]{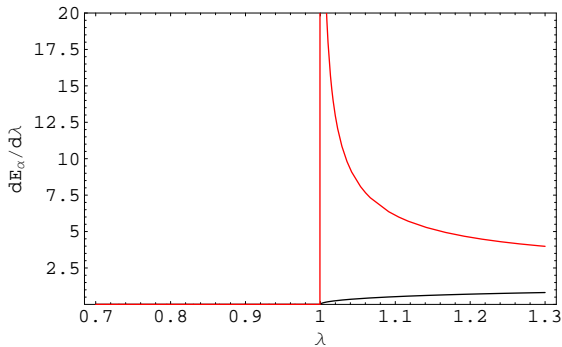}
\caption{First derivatives of the two independent energy gaps as a
function of $\lambda$ for the $XY$ model $(\gamma=0)$. Red line:
$d \Delta E_z /d \lambda$; black line: $d \Delta E_x/d \lambda = d
\Delta E_y/d \lambda$.} \label{f-derivative-3}
\end{figure}

Considering the isotropic $XY$ model
$(\gamma=0)$, the behavior of the energy gaps close to the
quantum critical point changes radically because the system
belongs to a universality class characterized by a new symmetry,
the invariance under rotations around the z-axis. This symmetry
imposes the plane degeneracy $E_x = E_y$ in \eqs{Energy Gap}. In
\fig{f-derivative-3} we show the behavior of the derivatives of
the two remaining independent energy gaps ($E_x$ and $E_z$) as a
function of $\lambda$ approaching the critical value $\lambda_c$.
One finds that the derivative of the energy gap $\Delta E_z$
associated to the $\pi$ phase gate operation $\sigma_{k}^{z}$
diverges as an inverse power approaching $\lambda_c$ from above:
\begin{equation}
\label{qptbehaviourxx} \frac{d \Delta E_z}{d \lambda} =
\frac{8}{\pi} (\lambda^2-\lambda_c^2)^{-1/2} \; .
\end{equation}
The derivative of the energy gap $\Delta E_x$ associated to the
spin flip gate remains finite for all values of $\lambda$. Hence,
only the energy gap associated to the ($\pi$) phase gate operation
detects the onset of the critical regime in the $XY$ model,
due to the larger symmetry constraints.

Next, we consider quantum state dynamics rather
than energy observables. In general, a local gate operation
perturbs the ground state $\ket{g}$, and under the subsequent
dynamics driven by \eq{Hamiltoniana} any observable that does not
commute with the system Hamiltonian acquires a nontrivial time
evolution. It is convenient to focus on the dynamics
of the magnetization along the $z$-axis on site $k$, because it is
nonvanishing for all values of $\lambda$, and hence its time
evolution is best suited for analytical treatment as well as for
experimental observation. Let us then define the dimensionless
time $\tau = (ht)/\hbar$, and consider the short-time dynamics
($\lambda \tau \ll 1$) of the magnetization on site $k$ after the
local gate operation has been performed at $\tau = 0$. One has $
<\sigma_k^z(\tau)> - <\sigma_k^z(0)> = - 4 \Lambda_\alpha (\lambda
\tau)^2$, where $<\sigma_k^z(0)>$ is the magnetization along the
$z$-axis evaluated on the perturbed state at time $0$,
$<\sigma_k^z(\tau)>$ is the on site magnetization at short times
$\tau$ after the action of the local gate, and $\Lambda_\alpha$ is
the short-time diffusion coefficient, or acceleration, whose form
depends on the single qubit operation that has been performed.
This acceleration can be expressed, similarly to
the case of the energy gaps previously studied, as a linear
combination of correlation functions evaluated in the ground
state. For the three fundamental operations ($\sigma_k^x$,
$\sigma_k^y$, and $\sigma_k^z$), the coefficients $\Lambda_\alpha$
can be worked out analytically and read, respectively,
\begin{eqnarray}
\label{evolutcoefficients}
\Lambda_x & = & - M_z + \frac{\gamma}{\lambda}(G_{xx}+G_{yy})- \nonumber \\
& & - \frac{1+\gamma}{2} G_{xzx}+\frac{\gamma(1-\gamma)}{2} G_{yzy} \; , \nonumber \\
\Lambda_y & = & -M_z - \frac{\gamma}{\lambda}(G_{xx}+G_{yy})- \\
& & -\frac{\gamma(1+\gamma)}{2} G_{xzx} - \frac{(1-\gamma)}{2} G_{yzy} \; , \nonumber \\
\Lambda_z & = & \gamma^2 M_z - \frac{\gamma}{\lambda}(G_{xx}-G_{yy}) \nonumber \\
& & + \frac{1+\gamma}{2}G_{xzx} + \frac{1-\gamma}{2}G_{yzy} \; .
\nonumber
\end{eqnarray}
Comparing \eqs{Energy Gap} and
\eqs{evolutcoefficients}, we see that in the latter there appear
three-point correlations involving nearest
neighbor and next to nearest neighbor sites: $G_{\alpha \beta
\alpha} = \bra{g} \sigma_{k-1}^\alpha \sigma_k^\beta
\sigma_{k+1}^\alpha \ket{g}$. The presence of correlations
of order $n>2$ is related to the dynamical nature of 
the observable. Including next-to-leading order powers of
$\lambda \tau$ in the expansion of 
$<\sigma_k^z(\tau)> - <\sigma_k^z(0)> $, 
one would need to evaluate four-point and higher-order
correlations.

\begin{figure}[t!]
\includegraphics[width=7.5cm]{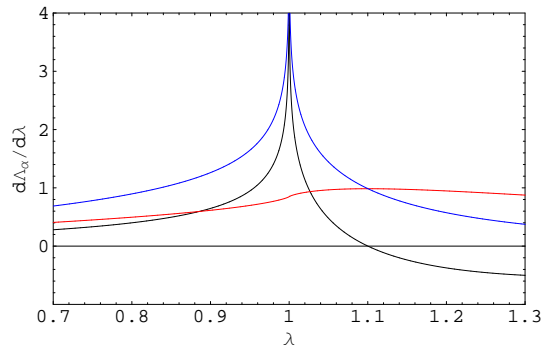}
\caption{First derivatives of the short-time diffusion
coefficients $\Lambda_x$ (black line), $-\Lambda_y$ (blue line),
and $\Lambda_z$ (red line), as functions of $\lambda$ for the
Ising model $(\gamma=1)$. Notice that $\Lambda_z$ is always
constant as a function of $\lambda$.} \label{der-velocity-3}
\end{figure}

In \fig{der-velocity-3} we plot the behavior of the derivatives of
the diffusion coefficients $ d \Lambda_\alpha /d \lambda$ ($\alpha
= x, y, z$) as functions of $\lambda$ for the Ising model
$(\gamma=1)$. The derivatives of $\Lambda_x$ and $\Lambda_y$
diverge logarithmically approaching $\lambda_c$, in analogy with
the derivatives of the energy gaps (see \eq{qptbehaviour} and
\fig{f-derivative}), as
\begin{equation}\label{qptbehaviourvelocity}
\frac{d \Lambda_\alpha}{d \lambda} = d_\alpha(\gamma) \ln |\lambda
- \lambda_c| + const \; .
\end{equation}
The coefficients $d_\alpha(\gamma)$ can be evaluated analytically.
In the Ising case the following symmetry holds: 
$d_x(1) = - c_x(1)$, $d_y(1) = - c_y(1)$. In the whole interval 
$0 < \gamma <1$ the accelerations $\Lambda_x$ and $\Lambda_y$ 
exhibit a very similar scaling behavior in the proximity 
of the quantum critical point, obeying in 
all cases \eq{qptbehaviourvelocity}. 
The coefficients $d_{x,y}(\gamma)$ in
\eq{qptbehaviourvelocity} are related to the coefficients
$c_{x,y}(\gamma)$ in \eq{qptbehaviour} by a simple translation of
minus the latter by an amount that decreases monotonically with
the value of the anisotropy and vanishes when the
latter reaches its maximum $\gamma=1$. As an example, at
$\gamma=0.5$ the coefficients $d_x(0.5)$ and $d_y(0.5)$ are
$d_x(0.5) = -c_x(0.5) + 0.239$ and $d_y(0.5)= -c_y(0.5) + 0.239$.

The reason why $d \Lambda_z / d \lambda$ does not exhibit
scaling in the Ising model can be understood by comparing
with its behavior at different values of the anisotropy. One finds
that $d \Lambda_z / d \lambda$ scales logarithmically, just
like $d \Lambda_x / d \lambda$ and $d \Lambda_y / d \lambda$, 
when approaching the quantum critical point in the whole  
interval $0 < \gamma < 1$. At first sight, this difference in the
behavior of the acceleration $\Lambda_z$ for the Ising model 
versus the other anisotropic cases would seem to imply a violation of 
the universality principle. However, studying the behavior of 
$\Lambda_z$ as a function of $\lambda$ when
the anisotropy parameter $\gamma$ approaches the Ising limit, 
one finds that $d \Lambda_z / d \lambda$ still obeys 
\eq{qptbehaviourvelocity}, but with a coefficient $d_z(\gamma)$ 
that decreases monotonically and vanishes exactly at $\gamma = 1$.

\begin{figure}[t!]
\includegraphics[width=7.5cm]{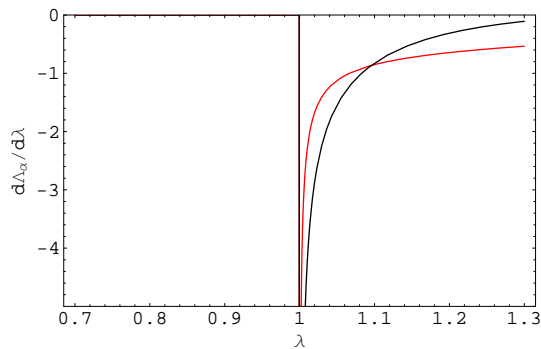}
\caption{Behavior of the first derivatives of the velocities
$\Lambda_x = \Lambda_y$ and of the velocity $\Lambda_z$ as
functions of $\lambda$ when approaching the critical point
$\lambda_{c}$ for the isotropic $XY$ model ($\gamma=0$). Red line:
$d \Lambda_{z}/d \lambda$; black line: $d \Lambda_{x}/d \lambda =
d \Lambda_{y}/d \lambda$.} \label{der-velocity-xx}
\end{figure}

In the case of the isotropic $XY$ model, we enter in
a different class of universality and the behavior of the
derivatives of the diffusion coefficients changes drastically
compared to the anisotropic cases. Looking at
\eq{evolutcoefficients} and taking into account that rotational
symmetry implies $G_{xzx} = G_{yzy}$, we obtain $\Lambda_x =
\Lambda_y = - G_z - G_{xzx}/2$. At variance with the behavior of
the derivatives of the three energy gaps (compare
\fig{f-derivative-3} and \fig{der-velocity-xx}), {\it all} the
derivatives of the three accelerations $\Lambda_x = \Lambda_y$ and
$\Lambda_z$ diverge when approaching the critical point with a
behavior analogous to that of the derivative of the energy gap
$\Delta E_z$ (see \eq{qptbehaviourxx}):
\begin{equation}
\label{qptbehaviourxxvelocity} \frac{d \Lambda_\alpha}{d \lambda}
= d_{\alpha}' (\lambda^2-\lambda_c^2)^{-1/2} \; ,
\end{equation}
with $d_{x}' = d_{y}' = - 1/\sqrt{2\pi^2}$ and $d_{z}' = -
\sqrt{2}/\pi$.

In conclusion, we have introduced and studied observable
quantities, either static or dynamic, associated with single qubit
unitary operations, that are able to characterize quantum phase
transitions in spin systems. This provides an alternative,
quantum-informatic point of view complementary to the traditional
one in terms of correlation functions. Moreover, at variance with
the characterization of quantum critical points based on measures
of entanglement, the quantities that we have introduced are, in
principle, amenable to {\it direct} experimental observation.
It is matter for future studies to establish whether this 
formalism can be useful for the characterization of other quantum 
critical systems and models, and in different research contexts.
In this last respect, it would be particularly interesting to investigate 
possible connections with quantum information transfer, information 
speed, and quantum speed limits in critical systems
\cite{Bose,Plenio,Giovannetti}.

We thank L. Amico, F. de Pasquale, and R. Fazio for useful discussions. 
Financial support from MIUR, Coherentia CNR-INFM, and INFN is acknwoledged.

\end{document}